\documentclass[12pt]{article}
\newcommand{\Dcslash}{/\!\!\!\! D}
\newcommand{\spartial}{/\!\!\!\partial}
\newcommand{\Aslash}{/\!\!\!\!A}
\newcommand{\Dslash}{/\!\!\!\!D}

\begin{document}
\title{Non local parity transformations and anomalies}
\author{M.~L.~Ciccolini$^a$~\footnote{Electronic address:
    ciccolin@ib.cnea.gov.ar}, C.~D.~Fosco$^a$~\footnote{Electronic
    address: fosco@cab.cnea.gov.ar} and
  F.~A.~Schaposnik$^b$~\footnote{Electronic address:
    fidel@athos.fisica.unlp.edu.ar}\\ \\
{\normalsize\it $^a$Centro At\' omico Bariloche - Instituto Balseiro,}\\ 
{\normalsize\it Comisi\'on Nacional de Energ\'{\i}a At\'omica}\\
{\normalsize\it 8400 Bariloche, Argentina}\\ \\
{\normalsize\it $^b$Departamento de F\'\i sica, Universidad Nacional
    de La Plata}\\
  {\normalsize\it C.C. 67, 1900, La Plata, Argentina} } \date{\hfill}
\maketitle
\begin{abstract}
\noindent
We present an alternative derivation of the parity anomaly for a
massless Dirac field in $2+1$ dimensions coupled to a gauge field. The
anomaly functional, a Chern-Simons action for the gauge field, is
obtained from the non-trivial Jacobian corresponding to a non local
symmetry of the Pauli-Villars regularized action. That Jacobian is
well-defined, finite, and yields the standard Chern-Simons term when
the cutoff tends to infinity.
\end{abstract}
\bigskip

\section{Introduction}
Because of the presence of ultraviolet divergences, most quantum field
theory models require the use of a regularization scheme at some
intermediate step in the renormalization program. The regularization
procedure, to be effective, will modify the large momentum behaviour
of the theory. In some cases, there are symmetries which are
particularly sensitive to that modification, leading to the existence
of anomalies, namely, symmetries which remain broken even after the
regularization is removed. In other words, performing a symmetry
transformation and renormalizing are non-commuting operations.

In the context of the functional integral quantization, anomalies are
traditionally attributed to the non invariance of the integration
measure under a classical symmetry transformation. The non-invariance
of the integration measure shows up in that the Jacobian corresponding
to the classical symmetry transformation is non trivial and, in 
general,ill-defined. When a proper regularization is introduced,
this Jacobian gives rise to an anomalous term
in the effective action~\cite{Fujikawa}.

In reference~\cite{fm}, an alternative procedure to study anomalies
within the functional integral approach was presented. It amounts to
considering the functional integral for a {\em regularized\/}
action, and to restrict the study to symmetry transformations that
leave that action invariant.  For the cases of the conformal and
chiral anomalies, we have shown that those symmetry transformations do
exist, and that they tend to the usual ones when the cutoff is
removed. The transformations are necessarily non local, but on a
length scale of the order of the inverse of the cutoff.  The outcome
of performing those symmetry transformations is that, while the
regularized classical action is indeed invariant (even for a finite
cutoff), the integration measure is not. Moreover, this non invariance
is explicit, since the corresponding Jacobian determinant is {\em well
  defined and different from $1$}, for any value of the cutoff.
Contact with the usual anomalies is made by taking the infinite cutoff
limit, what reproduces the known results.

In this letter, we will point out that this procedure can also be
applied to the case of the parity anomaly in $2+1$ dimensions, since
the regularized action {\it does\/} have a non local symmetry, which
coincides with the classical parity symmetry in the infinite cutoff limit.
Moreover, the associated Jacobian for this discrete transformation is
{\em finite}, and reproduces the parity anomaly, a Chern-Simons term in the
effective action
\cite{DeserJT}-\cite{RedlichL}. 
This shows that the procedure we have developed in
ref.~\cite{fm} can also be applied to the case of a {\em discrete\/}
symmetry.

The organization of this letter is as follows: In
section~\ref{sec:sym}, we discuss the classical parity transformations
and its generalization to the (quantum) regularized case. The
generalized transformations are then applied to the calculation of the
parity anomaly in section~\ref{sec:anom}.  Section~\ref{sec:concl}
contains our conclusions.

\section{Generalized parity transformations}\label{sec:sym}
Let us consider massive Dirac fermions coupled to an external Abelian
gauge field $A_{\mu}$, in $2+1$ dimensions. The (Minkowski spacetime)
action for the system is
\begin{equation}
S_f[A] \;=\; \int d^3x \, \mathcal{L}_f
\end{equation}
where the Lagrangian density $\mathcal{L}_f$ is given by
\begin{equation}
\mathcal{L}_f\,=\,\bar{\psi}(x)(i\,\spartial-e\,\Aslash(x)-m)\psi(x) \;.
\end{equation}
In our conventions, $g_{\mu\nu}={\rm diag}(1,-1,-1)$, and the
Dirac matrices verify
\mbox{$\{\gamma_{\mu},\gamma_{\nu}\}=2g_{\mu\nu}$}, and
$\gamma_{\mu}^\dagger=\gamma^{\mu}$.

We recall that, in $2+1$ dimensions, parity transformations are, in
fact, tantamount to spatial reflections:
\begin{equation}
\label{parity}
x=(x^0,x^1,x^2)  \stackrel{\mathcal{P}}{\longrightarrow}  x^P=(x^0,-x^1,x^2)
\end{equation}
since a spatial inversion is a rotation in $\pi$ when the number of
space coordinates is even.

The gauge and spinor fields transform according to the rules:
$$
A(x) =(A^0(x),A^1(x),A^2(x)) \, \to \,
A^P(x^P)=(A^0(x),-A^1(x),A^2(x))
$$
\begin{equation}
 \label{fieldsparity}
\psi(x) \, \to\, \psi^P(x^P)=\gamma_1\psi(x)
\;\;\; \bar{\psi}(x) \, \to \, \bar{\psi^P}(x^P) =
\bar{\psi}(x)\gamma_1 \;.
\end{equation}
Hence, under $\mathcal{P}$, the classical action transforms as
follows:
$$
S[A] \to S_f^P[A] \;=\; \int d^3x^P\,
\bar{\psi}_P(x_P)(i\,\spartial_P-e\,\Aslash_P(x_P)-m)\psi_P(x_P)
$$
\begin{equation}
=\;\int d^3x \, \bar{\psi}(x)(i\,\spartial-e\,\Aslash(x)+m)\psi(x) \;.
\end{equation}
Writing the mass dependence of the action explicitly, we have the
relation:
\begin{equation}
S_f^P(m)=S_f(-m) \;.
\end{equation}
This indicates, of course, that the massless classical theory is
parity invariant.

Regarding the regularized action, although not every regularization
method can be implemented in terms of an action, there are many
important cases where this can be done.  Examples of those are the
Euclidean cutoff, lattice, and Pauli-Villars (PV) regularization
methods. We shall use the PV method, since it maintains most of the
symmetries, except parity (see \cite{adm} for a thorough
discussion in odd-dimensional spaces).  This greatly simplifies the discussion.
In our $2+1$ dimensional example, the functional integral is rendered
convergent by the addition of just one bosonic regulator spinor
field $\phi$,
whose mass $\Lambda$ plays the role of a cutoff.  The regularized
action $S_f^r$ is:
\begin{equation}
\label{acreg}
S_f^r=i\int d^3x\left[\bar{\psi}(x)(\,\spartial+ie\,\Aslash )\psi(x) +
\overline{\phi}(x)(\,\spartial+ie\,\Aslash +i\Lambda)\phi(x)\right].
\end{equation}
It is possible to write the regularized action (\ref{acreg}) in terms
of just one fermionic field $\Psi$, although at the expense of
equipping that field with a non local action. To that end, we consider
the regularized functional integral $\mathcal{Z}^r[A]$ corresponding
to (\ref{acreg}):
\begin{equation}
\label{Z}
\mathcal{Z}^r[A]\;=\; \int\mathcal{D}\psi\mathcal{D}\bar{\psi}
\mathcal{D}\phi\mathcal{D}\bar{\phi} \,
\exp\left\{-\int d^3x
  [\bar{\psi}(x)\,\Dcslash\psi(x)+\bar{\phi}(x)
(\,\Dcslash+i\Lambda)\phi(x)] \right\}
\end{equation}
where $\Dcslash \equiv (\,\spartial+ie\,\Aslash )$.

We first integrate out the regulator in (\ref{Z}):
\begin{equation}
   \label{Zint}
\mathcal{Z}^r[A] \;=\; \int\mathcal{D}\psi\mathcal{D}\bar{\psi} 
\det^{-1}[1-\frac{i\,\Dcslash}{\Lambda}] \, 
\exp\left(-\int d^3x \bar{\psi}\, \Dcslash\psi\right)
\end{equation}
(we have neglected an irrelevant constant factor). We then make the
change of variables:
\begin{eqnarray}
\label{psitoPsi1}
\psi(x) &=&
\left[1-\frac{i\,\Dcslash}{\Lambda}\right]^{-\frac{1}{2}}
\;\Psi(x) \\
\label{psitoPsi2}
\bar{\psi}(x)&=&\bar{\Psi}(x)\;
\left[1-\frac{i\,\Dcslash}{\Lambda}\right]^{-\frac{1}{2}}
\end{eqnarray}
under which the measure transforms according to:
\begin{equation}
\mathcal{D}\psi\mathcal{D}\bar{\psi}=\mathcal{D}\Psi\mathcal{D}\bar{\Psi}
\det [1-\frac{i\,\Dcslash}{\Lambda}]\;.
\end{equation}
The result of the two previous steps is that we may rewrite
(\ref{Zint}) in the equivalent form:
\begin{equation}
\label{eq:snlp}
\mathcal{Z}^r[A]=\int\mathcal{D}\Psi\mathcal{D}\bar{\Psi} 
\exp(- S_f^{nl})
\end{equation}
where $S_f^{nl}$ denotes a non-local form of the regularized action,
\begin{equation}
  \label{eq:snl}
  S_f^{nl}\;=\; \int d^3x\,d^3y\, \bar{\Psi}(x) \mathcal{D}(x,y) \Psi(y) \;.
\label{efact}
\end{equation}
with
\begin{equation}
{\mathcal
  D}(x,y)=\left[\frac{\Dcslash}{1-\displaystyle{\frac{i\,\Dcslash}{\Lambda}}}\right]
(x,y) \;.
\end{equation}
We have found it convenient to adopt a `bracket' like notation to
write the action (\ref{eq:snl}),
\begin{equation}
  \label{eq:snl1}
  S_f^{nl}\;=\; \langle\Psi|\mathbf{D}|\Psi\rangle \;\;,\;\; 
{\mathbf D}\;=\; \frac{\Dcslash}{1-\displaystyle{\frac{i\,\Dcslash}{\Lambda}}}\,,
\end{equation}
since it avoids writing operators kernels and integrations explicitly.
Note that the `bra' includes the $\gamma_0$ factor for the adjoint
field.

To understand the behaviour of (\ref{eq:snl}) under parity
transformations, we previously need to know the ${\mathbf D}$ operator
transformation properties. Denoting by $\Dcslash(x,y)$ the kernel of
$\Dcslash$, we see that
\begin{equation}
\gamma_1 \,\Dcslash^P(x^P,y^P) \,\gamma_1 \;=\; \Dcslash(x,y) \;.
\end{equation}
We can also express this result as:
\begin{equation}
\label{Dcparity}
\gamma_1\, \Dcslash^P \,\gamma_1 \;=\;\Dcslash \;,
\end{equation}
what in turn implies for ${\mathbf D}$:
\begin{equation}
\label{DParity}
\gamma_1{\mathbf D}^P(\Lambda)\gamma_1 \;=\; {\mathbf D(-\Lambda)}\,.
\end{equation}

The parity transformed non local regularized action becomes
\begin{equation}
S_f^{nlP}=\langle\Psi^P| {\mathbf D}^P(\Lambda) |\Psi^P\rangle = 
\langle\Psi^P| \gamma_1 {\mathbf D}(-\Lambda) \gamma_1|\Psi^P\rangle \;,
\end{equation}
where we need to plug in the parity transformed of $\Psi$ and
${\bar\Psi}$. The transformation rules for the new fields, can be
obtained as follows: from the relation between $\Psi$ and $\psi$, we
learn that
\begin{equation}
\psi^P(x^P) \;=\;
\left[1-\frac{i\,\Dcslash^P}{\Lambda}\right]^{-\frac{1}{2}}\,
\Psi^P(x^P)
\end{equation} 
and
\begin{equation}
\psi^P(x^P)=-\gamma_1 \left[1+\frac{i\,\Dcslash}{\Lambda}\right]^{-\frac{1}{2}}
\gamma_1\Psi^P(x^P) \;.
\end{equation}
On the other hand, we have the relation
\begin{equation}
\psi^P(x^P) = \gamma_1 \psi (x) \;=\;
\gamma_1 \,\left(1-\frac{i\,\Dcslash}{\Lambda}\right)^{-\frac{1}{2}}\Psi(x) \;.
\end{equation}
Then,
\begin{equation}
\Psi^P(x^P)=\gamma_1\frac{\sqrt{1+\frac{i\,\Dcslash}{\Lambda}}}
{\sqrt{1-\frac{i\,\Dcslash}{\Lambda}}}\Psi(x)=\gamma_1\langle
x|\frac{{\mathbf D}(\Lambda)}
{\sqrt{{\mathbf D}(\Lambda){\mathbf D}(-\Lambda)}} |\Psi \rangle \;.
\end{equation}

We obtain an analogous relation for $\bar{\Psi}(x)$. Using the compact
notation, we see that

\begin{equation}
 \label{eq:comp1}
|\Psi^P\rangle=\gamma_1 \frac{{\mathbf D}(\Lambda)}{\sqrt{{\mathbf D}(\Lambda){\mathbf D}(-\Lambda)}} |\Psi\rangle
\end{equation}
\begin{equation}
 \label{eq:comp2}
\langle \Psi^P|=\langle\Psi|\frac{{\mathbf D}(\Lambda)}{\sqrt{{\mathbf
      D}(\Lambda){\mathbf D}(-\Lambda)}} \gamma_1 \;,
\end{equation}
and these are the generalized parity transformations we were looking
for. We note that they tend to the standard parity transformations
when the cutoff is removed ($\Lambda \to \infty$).

It is immediate to verify that, with these transformation rules, the
non-local form of the regularized action remains invariant:
\begin{equation}
 \label{eq:snlinv}
S_f^{nl P}=S_f^{nl} \;.
\end{equation}
Being the Pauli-Villars fields integrated out exactly,
the effective action (\ref{efact})
 incorporates the complete effect of the regularization, in
the sense that 
 the contribution to $S_f^{nl}$ of the
determinant associated with the Pauli-Villars field is totally
 included. For 
example, were one to use this effective action to compute the
vacuum polarization, the corresponding diagrams will automatically
be convergent to all orders. 

\section{Parity anomaly}\label{sec:anom}

Although the regularized action is invariant under the generalized
parity transformations, the fermionic measure acquires a non trivial
Jacobian:
\begin{equation}
\mathcal{D}\Psi_P\mathcal{D}\bar{\Psi}_P =
\mathcal{D}\Psi\mathcal{D}\bar{\Psi}\; {\mathcal J}
\end{equation}
where
\begin{equation}
{\mathcal J} \;=\;\det\left[-\left(\frac{\sqrt{D(\Lambda)D(-\Lambda)}}{D(\Lambda)}
  \right)^2\right] = \det\left(
  \frac{i\,\Dcslash -\Lambda}{i\,\Dcslash +\Lambda} \right) \;.
\end{equation}

The parity anomaly is obtained from the infinite-$\Lambda$ limit of
this Jacobian. This result is known, and coincides of course with
(twice) the leading term in a derivative expansion of the effective
action $I_f^{\mathrm{eff}}[A,\Lambda]$: We see that:
\begin{equation}
\exp(iI_{\mathrm{eff}}[A,\Lambda]\,-\,iI_{\mathrm{eff}}[A,-\Lambda])
\;=\; \det\left( \frac{\Dcslash+i\Lambda}{\Dcslash-i\Lambda}\right) \;=\; \mathcal{J}\;.
\end{equation}
\begin{equation}
I_{\mathrm{eff}}[A,\Lambda] = -i \ln \det(\,\Dcslash+i\Lambda).
\end{equation}
Hence the Jacobian is expressed as a function af the effective action.
In the $\Lambda\to\infty$ one obtains~\cite{Dunne}-\cite{S}:
\begin{equation}
I_{\mathrm{eff}}[A,\Lambda]\;\to\; \frac{\Lambda}{|\Lambda|} S_{C-S}\;,
\end{equation}
where the $C-S$ action $S_{C-S}$ is defined by:
\begin{equation}
  \label{eq:scs}
  S_{C-S}\;=\;\frac{e^2}{4\pi} \int d^3x\,\epsilon^{\mu\nu\rho}
  A_{\mu}(x) \partial_{\nu} A_{\rho}(x) \;.
\end{equation}
Thus we can write for the Jacobian:
\begin{equation}
\label{eq:jac}
\mathcal{J} \;=\; \exp\left\{ \pm 2 i S_{C-S}[A] \right\} \;, 
\end{equation}
which is the parity anomaly result. It is not renormalized by higher
order correction, as a consequence of the Coleman-Hill
theorem~\cite{Coleman-Hill}.

For the case of a continuous symmetry transformation, the Fujikawa
Jacobian resulting from the associated infinitesimal 
change of variables
can be used to calculate the potentially anomalous divergence of the corresponding
Noether current. This procedure does not, of course, have a direct
parallel in the present, discrete symmetry case.  Nevertheless, from
the knowledge of the Jacobian (\ref{eq:jac}), one can make explicit
the anomalous behavior of the fermion current under a parity
transformation. Indeed, from eq.(\ref{eq:snlp}), we can write
\begin{equation}
\mathcal{Z}^r[A^P]\equiv \exp\left(-I_{eff}^r[A^P]\right)
=\int\mathcal{D}\Psi\mathcal{D}\bar{\Psi} 
\exp(- S_f^{nl}[A^P,\bar \Psi, \Psi]) \;.
\label{ap}
\end{equation}
Now, changing the fermion variables as in
(\ref{eq:comp1})-(\ref{eq:comp2}), and using the invariance
(\ref{eq:snlinv}) of the non local action, we see that
\begin{equation}
\mathcal{Z}^r[A^P]={\cal J}[A^P]\int\mathcal{D}\Psi^P\mathcal{D}\bar{\Psi}^P 
\exp(- S_f^{nl}[A^P,\bar \Psi^P, \Psi^P]) = {\mathcal J}[A^P]
\mathcal{Z}^r[A]
\label{apf}
\end{equation}
or
\begin{equation}
\exp\left(-I_{eff}^r[A^P] + I_{eff}^r[A]\right)) = {\cal J}[A^P] \;.
\label{ac2}
\end{equation}
Differentiation of the regularized effective action with respect
to the gauge field $A$ yields the ground state current $\langle
j_\mu[A] \rangle$ so that one has
\begin{equation}
-e\langle j_\mu[A^P] \rangle + e\langle j_\mu[A] \rangle= 
\frac{\delta \log {\cal J}[A^P]}{\delta A^\mu} \;.
\end{equation}
Using relations (\ref{fieldsparity}) for the l.h.s.\ and (\ref{eq:jac})
for the r.h.s., we can then write the anomalous, parity-odd vacuum
current:
\begin{equation}
\langle j_\mu [A] \rangle^{odd} \;=\; \frac{1}{2e}  
\frac{\delta \log {\cal J}[A^P]}{\delta A^\mu} =
\pm i\frac{e}{4\pi}
\int d^3x\,\epsilon^{\mu\nu\rho}
  \partial_{\nu} A_{\rho}(x) \;.
\end{equation}
which is the well-known result first obtained in 
\cite{RedlichL}.

It is immediate to verify that all the steps leading the parity
anomaly for the Abelian case can be generalized to the non-Abelian
case as well, with the only modification that the Jacobian is now
related to the non-Abelian massive determinant.  The result is of
course that this Jacobian is the exponential of ($2 i$ times) the
non-Abelian Chern-Simons action.

\section{Conclusions}\label{sec:concl}

We have shown that the parity anomaly in 2+1 dimensions in a theory of
massless fermions coupled to an external gauge field can be obtained
from the Jacobian for a generalized symmetry transformation of the
regularized theory. This procedure has the virtue of disentangling the
symmetries and infinities, by looking for transformations which are
symmetries of the {\em regularized\/} action. This avoids involving
modes with an arbitrarily high momentum into the symmetry
transformation, and the bonus is that the resulting Jacobian
automatically takes care of the anomaly. It is worth noting that it is
precisely the fact that the regularization breaks the symmetry what
renders the symmetry transformations of the regularized action
non-local.  Non-anomalous local symmetries have, in this setting, a
distinctive property: they are always local, even when acting on the
regularized action.

Our calculations required the use of a non-local form of the
regularized action and we used `regularized' fermion fields $\Psi$.
The non-locality of the action, however, is hidden in the non-physical
region of momenta above the cutoff. For example, when $A=0$, the
${\mathbf D}$ operator has a kernel:
\begin{equation}
\Dslash(x,y) = \,\spartial \left( 1+\frac{i\,\spartial}{\Lambda}
\right) \int \frac{d^3k}{(2\pi)^3} \;
\frac{e^{-ik(x-y)}}{1-(k/\Lambda)^2}
\end{equation}
which is non-local on a scale ${1}/{\Lambda}$.

\section*{Acknowledgements}
C.~D.~F. would like to acknwledge Prof.~F.~D.~Mazzitelli, for useful
conversations and correspondence.

\noindent M.~L.~C.~is supported by a CNEA fellowship.  C.~D.~F.~is a member
of CONICET, and is partially supported by ANPCyT (PICT 97/1040) and
Fundaci\'on Antorchas grants.  F.~A.~S~is partially supported by CICBA
as Investigador, and through grants CONICET (PIP 4330/96), and ANPCyT
(PICT 97/2285).

\newpage


\begin{thebibliography}{bib}
\bibitem{Fujikawa}K.~Fujikawa, Phys.~Rev.~Lett. {\bf 42}, 1195 (1979);
 Phys.~Rev.~{\bf D21}, 2848 (1980).
\bibitem{fm}C.~D.~Fosco and F.~D.~Mazzitelli, Phys.~Lett.~{\bf B481},
  129 (2000). 
\bibitem{DeserJT}S.~Deser, R.~Jackiw and S.~Templeton, 
Phys. Rev. Lett. {\bf 48} (1982) 975;
Ann.~of~Phys. {\bf 140}, 372 (1982).  
\bibitem{RedlichL}A.~N.~Redlich, Phys.~Rev.~Lett.~{\bf 52}, 18 (1984); Phys.~Rev.~{\bf D29}, 2366 (1984).
\bibitem{adm} L.~Alvarez-Gaum\'e, S.~Della Pietra and G.~Moore, Ann. of
Phys.  (N.Y.) {\bf 163} (1985) 288.
\bibitem{Dunne}For a review, see:
  G.~V.~Dunne, Lectures at the 1998 Les Houches Summer School: {\em
    Topological Aspects of Low Dimensional Systems}, e-print arXiv:
  hep-th/9902115.  
\bibitem{S}R.~E.~Gamboa Sarav\'{\i},
  G.~L.~Rossini and F.~A.~Schaposnik, Int.~J.~Mod.~Phys. {\bf A11},
  2643 (1996).  
\bibitem{Coleman-Hill}S.~Coleman and B.~Hill,
  Phys.~Lett.~{\bf B159}, 184 (1985).
\end{thebibliography}
\end{document}